\documentstyle[12pt]{article}
\def\be{\begin{equation}}
\def\ee{\end{equation}}
\def\bea{\begin{eqnarray}}
\def\eea{\end{eqnarray}}
\def\ed{\end{document}}
\def\bd{
\begin{document}}
\def\bit{\begin{itemize}}
\def\eit{\end{itemize}}


\def\sig{\sigma}
\def\Sig{\Sigma}
\def\lam{\lambda}
\def\Lam{\Lambda}
\def\Del{\Delta}
\def\del{\delta}
\def\Bg{\Bar g}
\def\hg{\hat g}
\def\k{\kappa}
\def\alf{\alpha}
\def\ga{\gamma}
\def\Ga{\Gamma}
\def\BD{\Bar D}

\def\Lix{\pounds_\xi}
\def\di{\partial}
\def\nab{\nabla}
\def\half{{\textstyle{1 \over 2}}}
\def\~{\tilde}
\def\lag{\hat{\cal L}}
\def\m{\label}
\def\l{\left}
\def\r{\right}
\def\goto{\rightarrow}
\def\Bar{\overline}
\def\const{\rm const}

\bd

\centerline{\bf CONSERVED CURRENTS IN $D$-DIMENSIONAL
GRAVITY}
 \centerline{\bf  AND BRANE COSMOLOGY\footnote{Firstly,
 a more part of the results were presented on the Conference
 GR16 \cite{Petrov}.}}

\vskip .1 in
 \centerline{Alexander N. Petrov\footnote{E-mail:
anpetrov@rol.ru}}

\centerline{Stergnberg Astronimical Institute,}
\centerline{ Universitetskii
prospect 13, Moscow 119992, Russia}

\vskip .3 in
\begin{quote}
{\small \it In $D$-dimensional gravity on arbitrary curved
backgrounds using proven methods conserved currents, divergences of
antisymmetric tensor densities (superpotentials), are constructed.
The superpotentials have two remarkable properties: they depend in
an essential way on second derivatives in the Lagrangian and are
independent on divergences added to it. In particular, these
conserved currents are well adapted  to the case of perturbations in
Gauss-Bonnet cosmological brane theories.}
\end{quote}
\bigskip

\noindent In general relativity (GR),  gauge theories and
supergravity, as a rule, conserved currents (vector densities) are
constructed as  divergences of antisymmetrical tensor densities
(superpotentials). Frequently the use of an auxiliary background
spacetime turns out even inevitable (see \cite{[1]}~-~\cite{[7]} and
references there in). With this one has to note that there are other
approaches without using backgrounds. One of them is developed,
e.g., in \cite{ACOTZ, ACOTZ1}, where the N{\oe}ther charges
connected with asymptotic Killing vectors in asymptotically anti-de
Sitter theories of gravity are constructed with using special
surface terms added to a Lagrangian.

Here, developing the approach with backgrounds we consider an
arbitrary $D$-dimensional theory of gravity. Keeping in mind, say,
scalar-tensor theories of gravity, or other generalizations of
gravity, we use the generalized gravitational variable $A_B \in
g_{\mu\nu},\, \Phi_B$ where $g_{\mu\nu}$ is $D$-dimensional metric
and $\Phi_B$ is a set of tensor density fields (not spinor ones).
(If it is necessary, it could be assumed that $\Phi_B$ includes
matter, non-gravitational variables.) We assume that Lagrangian
density $\hat L(A_B;\,A_{B,\alf};\,A_{B,\alf\beta})$ for
gravitational field includes the second derivatives as well as the
first ones; hats ``$~\hat {}~$'' mean densities of the weight $+1$.
We consider also an external auxiliary metric $\Bar g_{\mu\nu}$ of a
fixed arbitrary curved background spacetime with the Riemannian
tensor $\Bar R^\alf_{~\beta\mu\nu}$; bars mean background
quantities. To include $\Bar g_{\mu\nu}$ we, as usual \cite{[8]},
rewrite the ordinary derivatives $({}_{,\alf})$ over the  covariant
with respect to $\Bar g_{\mu\nu}$ derivatives $({}_{;\alf})$. Then
the Lagrangian acquires an explicitly covariant form:
 \be \hat L =
\lag = {\lag} (A_B;\, A_{B;\alf}; \,
   A_{B;\alf\beta}).
 \m{(1)}
 \ee
Now indexes  are shifted by $\Bar g_{\mu\nu}$ and $\Bar g^{\mu\nu}$.

The Lagrangian (1), as
a scalar density,  satisfies
the identity:
$
 \pounds_\xi \lag + (\xi^\alf \lag){}_{,\alf} \equiv 0$
where, following to notations in \cite{[8]},
we define the Lie derivative with respect to a vector field
$\xi^\alf$ as
 $ \pounds_\xi A_B \equiv
 - \xi^\alf  A_{B;\alf} +  \xi^\beta{}_{;\alf} \l.A_B\r|^\alf_\beta
 $. This identity can be rewritten in the form:
 \be
\l[L^B
  A_{B;\alf} +  (L^B
\l. A_B\r|^\beta_\alf){}_{;\beta}\r]\xi^\alf
-  \l[\hat u^{\alf}_\sig
 \xi^\sig + \hat
 m^{\alf\beta}_{\sig} \xi^\sig{}_{;\beta} + \hat
 n^{\alf\beta\gamma}_\sig \xi^\sig{}_{;\beta\gamma}\r]_{;\alf}
 \equiv 0.
 \m{(2)}
 \ee
In (\ref{(2)}), the coefficients are defined by the Lagrangian in unique
way:
\be
 \hat u^\alf_\sig  \equiv \lag \delta^\alf_\sig +
 L^B
 \l.A_B\r|^\alf_\sig  -
 M^{B\alpha }
 A_{B;\sig}  -
 N^{B\alpha \beta}
 A_{B;\beta\sig} - \hat
n^{\alf\beta\gamma}_\lam \Bar R^\lam_{~\beta\gamma\sig},
 \m{(3)}
\ee
\be
\hat m^{\alf\beta}_\sig  \equiv
M^{B\alpha}
 \l.A_{B}\r|^\beta_\sig +
 N^{B\alpha\gamma}
 \l[  ( \l.A_{B}\r|^\beta_\sig)_{;\gamma} -\delta^\beta_\gamma
 A_{B;\sig} \r],
 \m{(4)}
\ee
\be \hat n^{\alf\beta\gamma}_\sig \equiv
 N^{B\alpha(\beta}
 \l.A_{B}\r|^{\gamma)}_\sig.
 \m{(5)}
\ee We use  the notations for Lagrangian derivatives $L^B \equiv
\delta \lag/\delta A_B$, and
$$
M^{B\alpha } \equiv
{{\di \lag} /
{\di  A_{B;\alf}}} -
  \l({{\di \lag} /
  {\di  A_{B;\alf\beta}}}\r){}_{;\beta},
~~~ N^{B\alpha\beta} \equiv
{{\di \lag} /
 {\di  A_{B;\alf\beta}}}.
 $$
The coefficient at $\xi^\sig$ in the first term in  (\ref{(2)})
is identically equal to zero (generalized Bianchi identity).
The identity (\ref{(2)}) has thus the
form of the
differential conservation law:
\be
 \hat I^\alf{}_{;\alf} \equiv  \hat
 I^\alf{}_{,\alf} \equiv 0.
 \m{(6)}
\ee For a generalized current we prefer to use  the helpful form:
\be
 \hat
 I^\alf \equiv - \l[(\hat u^{\alf}_\sig + \hat
n^{\alf\beta\gamma}_\lam \Bar R^\lam_{~\beta\gamma\sig})\xi^\sig + \hat
 m^{\alf\beta}_{\sig}\bar g^{\sig\rho} \xi_{[\rho,\beta]} +
 \hat z^{\alf}(\xi)\r]
 \m{(7)}
 \ee
with $z$-term  defined as
 \be 2 \hat z^{\alf}(\xi) \equiv  - \hat
m^{\alf\beta}_{\sig}\bar g^{\sig\rho}{\pounds}_\xi \Bar
g_{\beta\rho}+ \hat n^{\alf\beta\gamma}_\sig \Bar g^{\sig\rho}
\l[({\pounds}_\xi \Bar g_{\beta\gamma})_{;\rho} - 2 ({\pounds}_\xi
\Bar g_{\beta\rho})_{;\gamma}\r]. \m {(8)}
 \ee
If $\xi^\alf$ is a Killing vector of the background, then $\hat
z^{\alf}(\xi) = 0$ and the current (\ref{(7)}) is determined by the
energy-momentum $u$-term and the spin $m$-term. Opening the identity
(\ref{(6)}) and, since $\xi^\sig$,  $ \xi^\sig{}_{,\alf}$, $
\xi^\sig{}_{,\alf,\beta}$ and $ \xi^\sig{}_{,\alf,\beta,\gamma}$ are
arbitrary at every world point, equalizing independently to zero the
coefficients at $\xi^\sig$,  $ \xi^\sig{}_{;\alf}$, $
\xi^\sig{}_{(;\alf;\beta)}$ and $ \xi^\sig{}_{(;\alf;\beta;\gamma)}$
we get a ``cascade'' (in terminology by Julia and Silva \cite{[2]})
of identities: \bea
 &{}&   \hat u^{\alf}_{\sig;\alf} + \half
 \hat m^{\alf\rho}_\lam \Bar R^{~\lam}_{\sig~\rho\alf}
 +{\textstyle{1\over 3}} \hat n^{\alf\rho\gamma}_\lam
 \Bar R^{~\lam}_{\sig~\rho\alf;\gamma}
  \equiv 0, \nonumber \\
&{}&    \hat u^{\alf}_\sig +
\hat m^{\lam \alf}_{\sig~;\lam} + \hat n^{\tau\alf\rho}_\lam
 \Bar R^{~\lam}_{\sig~\rho\tau} +{\textstyle{2\over 3}} \hat
 n^{\lam\tau\rho}_\sig \Bar R^{\alf}_{~\tau\rho\lam} \equiv 0,
 \nonumber \\ &{}&
 \hat m^{(\alf\beta)}_\sig +
 \hat n^{\lam(\alf\beta)}_{\sig~~~~;\lam} \equiv 0, \nonumber\\ &{}&
 \hat
 n^{(\alf\beta\gamma)}_\sig \equiv 0.
 \m{(9)}
\eea

On the other hand, since Eq. (\ref{(6)}) is identically satisfied,
the current (7) must be a divergence of a superpotential
$\hat \Phi^{\alf\beta}$ for which $ \hat \Phi^{\alf\beta}{}_{,\alf\beta}
\equiv 0$, that is
\be
 \hat I^\alf \equiv  \hat \Phi^{\alf\beta}{}_{;\beta} \equiv
   \hat \Phi^{\alf\beta}{}_{,\beta}.
 \m{(10)}
\ee
Indeed,  substituting $\hat u^\alf_\sig$ and
$\hat m^{\alf\beta}_\sig$ from Eqs.~(\ref{(9)})
{\it directly} into the current and using
algebraic properties of $n^{\alf\beta\gamma}_\sig$ and
$\Bar R^{\alf}_{~\beta\rho\sig}$
we  reconstruct (\ref{(7)}) into the form (\ref{(10)})
where  the superpotential is
\be
 \hat \Phi^{\alf\beta}  \equiv \l(\hat
 m^{\beta\alf}_\sig +  \hat
 n^{\lam\beta\alf}_{\sig~~~;\lam}\r)\xi^\sig + {\textstyle{2\over 3}}
 \xi^\sig \hat n^{[\alf\beta]\lam}_{\sig~~~~;\lam}  -
 {\textstyle{4\over 3}} \hat n^{[\alf\beta]\lam}_\sig
 \xi^\sig{}_{;\lam}.
 \m {(11)}
 \ee
It is antisymmetric in $\alf$ and $\beta$ by the third identity in
Eqs.~(\ref{(9)}). Of course,  Eq. (\ref{(10)}) is only the other
form of the differential conservation law Eq. (\ref{(6)}).

Let us find contributions into currents and superpotentials from a
divergence in the Lagrangian $\delta \lag =  \hat k^\nu{}_{,\nu}$.
For the scalar density $ \hat k^\nu{}_{,\nu}$ one has the identity $
(\Lix \hat k^\alf + \xi^\alf
 \hat k^\nu{}_{,\nu})_{,\alf} \equiv 0 $ which gives contributions (i)
$\delta \hat I^\alf$ into the current (\ref{(7)}) with
coefficients:
 $
 \delta\hat u^\alf_\sig  = 2
(\delta^{[\alf}_\sig \hat k^{\beta]} ){}_{;\beta}$,
$ \delta\hat m^{\alf\beta}_\sig =
 2 \delta^{[\alf}_\sig \hat k^{\beta]} $,~~
$ \delta\hat n^{\alf\beta\gamma}_\sig = 0$;  and (ii)
$ \delta\hat \Phi^{\alf\beta}  = -2 \xi^{[\alf} \hat
 k^{\beta]} $ into the
superpotential (\ref{(11)}). Then, for $\lag \goto \lag +
 \delta\lag$  Eq. (\ref{(10)}) changes as
\be
 \hat I^\alf+ \delta \hat I^\alf \equiv
  \l(\hat \Phi^{\alf\beta} + \delta \hat
 \Phi^{\alf\beta}\r){}_{;\beta},
 \m{(12)}
\ee
where the changes do not depend on a structure of
$\hat k^{\nu}$.

It is well known that without changing the identity (\ref{(6)}) we
can add to the current an arbitrary quantity $\Delta \hat
I^\alf(\xi)$ satisfying  $ [\Delta \hat I^\alf(\xi)]_{,\alf} \equiv
0 $. Analogously, without changing   $\hat I^\alf$ in (\ref{(10)})
the superpotential can be added by $\Delta \hat
\Phi^{\alf\tau}(\xi)$, if $[\Delta \hat
\Phi^{\alf\beta}(\xi)]_{,\beta} \equiv 0 $.  However, the ``broken''
current and superpotential can be ``improved'' by the same way
because the quantities  $\Delta \hat I^\alf(\xi)$ and $\Delta \hat
\Phi^{\alf\beta}(\xi)$ are not connected {\it at all} with the above
procedure applied to the given Lagrangian. Whereas, since the
current and the superpotential in Eq. (\ref{(10)}) are defined by
the coefficients (\ref{(3)}) - (\ref{(5)}), they are {\it uniquely}
determined by the Lagrangian and the outlined procedure. This claim
develops also the criteria by Szabados \cite{[9]} who suggested to
consider a connection of pseudotensors with Lagrangians ``as a
selection rule to choose from the mathematically possible
pseudotensors''.

Next, using an old rule due to Belinfante (see \cite{[10]} and
\cite{[7]}) we define a tensor density \be \hat s^{\alf\beta\sig}
\equiv - \hat s^{\beta\alf\sig} \equiv -
 \hat m^{\sig[\alf}_\lam \bar g^{\beta]\lam} -
 \hat m^{\alf[\sig}_\lam \bar g^{\beta]\lam} + \hat m^{\beta[\sig}_\lam
 \bar g^{\alf]\lam}
 \m{(13)}
 \ee
and add  $
(\hat s^{\alf\beta\sig}\xi_{\sig})_{;\beta}$ to both sides of
(\ref{(10)}).
This generates a new  conservation law
of the form:
 \be
 \hat I^\alf_{(b)} \equiv  \hat \Phi^{\alf\beta}_{(b)}{}_{;\beta}
 \equiv  \hat \Phi^{\alf\beta}_{(b)}{}_{,\beta}.
 \m{(14)}
 \ee
This modification cancels the spin
from the current (\ref{(7)}):
\be
\hat I^\alf_{(b)} \equiv
\l(- \hat u^{\alf}_\sig  +
 \hat s^{\alf\beta}{}_{\sig;\beta}\r)  \xi^\sig+
 \hat z^{\alf}_{(b)}(\xi)  \equiv
\hat u_{(b)\sig}^\alf  \xi^\sig+
 \hat z^{\alf}_{(b)}(\xi),
 \m{(15)}
 \ee
and a new $z$-term appears, which is also equal to zero for Killing
vectors of the background. Thus, the  current $\hat I^\alf_{(b)}$ is
defined, in fact, by the modified energy-momentum tensor density
 $
 \hat u_{(b)\sig}^\alf$.
Because the new superpotential depends on  the $n$-coefficients
  only:
 \be
 \hat
 \Phi^{\alf\beta}_{(b)}  \equiv 2
\l({\textstyle{1\over 3}} \hat n^{[\alf\beta]\rho}_{\sig~~~~;\rho}+ \hat
 n^{\tau\rho[\alf}_{\lam~~~~;\tau} \bar g^{\beta]\lam} \bar g_{\rho\sig}\r)
\xi^\sig - {\textstyle{4\over 3}} \hat
 n^{[\alf\beta]\lam}_\sig\xi^\sig{}_{;\lam},
 \m{(16)}
\ee
it does not exist for Lagrangians with only the first
order derivatives. On the other hand,
the superpotential (16) is well adapted to theories
with second derivatives in Lagrangians,
like GR or brane cosmologies with Gauss-Bonnet Lagrangian.
With (\ref{(13)})
we conclude that  in the sense of our procedure the
current $\hat I^\alf_{(b)}$ and the superpotential $\hat
\Phi^{\alf\beta}_{(b)}$
are defined also in unique way by the Lagrangian (\ref{(1)}).

It is important to  note that the new currents (15) and superpotentials (16)
are independent on divergences $ \delta \lag = \hat k^\nu{}_{,\nu}$ added
to the Lagrangian (see also \cite{[7]}).
Indeed, the quantity
(\ref{(13)}) constructed for
$\delta \hat m^{\alf\beta}_{\sig}$ (see (\ref{(12)})) gives
 $
 \delta \hat s^{\alf\beta\sig}\xi_\sig =
 2\xi^{[\alf}\hat k^{\beta]}
 $.
The addition of
$(\hat s^{\alf\beta\sig}\xi_{\sig}
+ \delta \hat s^{\alf\beta\sig}\xi_{\sig})_{;\beta}$
to Eq. (\ref{(12)}) cancels the full spin m-term and suppresses
$\delta \hat u^{\alf}_\sig$ and
$\delta \hat \Phi^{\alf\beta}$
 because
 $
\delta \hat u^{\alf}_\sig - (\delta \hat
s^{\alf\beta\rho}\bar g_{\rho\sig})_{;\beta} =  0$~and~ $
\delta \hat \Phi^{\alf\beta} + \delta \hat
s^{\alf\beta\sig}\xi_\sig = 0$ for an arbitrary
$\hat k^\nu$.

Notice also that  conservation laws in the form
(\ref{(10)}), (\ref{(12)}) or (\ref{(14)}) are identities.
To transform them into
physically sensible conservation laws  one has to use field
equations in the currents on the left hand sides.

As an example, consider a perturbed model of a brane cosmology.
Let the role of $A_B$ in (\ref{(1)})
is only played by the physical metric $g_{\mu\nu}$
of a 5-dimensional bulk;
$g = {\rm det}\,g_{\mu\nu}$.
The physical curvature tensor with including
the background metric $\Bar g_{\mu\nu}$ can be rewritten in the form:
\be
 R^\lam{}_{\tau\rho\sig}  =
 \Delta^\lam_{\tau\sig;\rho} -  \Delta^\lam_{\tau\rho;\sig} +
 \Delta^\lam_{\rho\eta} \Delta^\eta_{\tau\sig} -
 \Delta^\lam_{\eta\sig} \Delta^\eta_{\tau\rho}
 + \Bar R^\lam{}_{\tau\rho\sig}
 \m{(17)}
\ee
(see, e.g., \cite{[1],[7]}), where the tensor $ 2\Delta^\alf_{\mu\nu}
\equiv g^{\alf\rho}(g_{\rho\mu;\nu} + g_{\rho\nu;\mu} -
g_{\mu\nu;\rho})$.
The Einstein-Gauss-Bonnet Lagrangian
is
\cite{[11], [12]}:
\be
\lag^{(5)} = -\half M_*^3\sqrt{-g}[R - 2\Lambda  +
\alf\underbrace{\l(R^2 - 4 R^{\alf\beta}
R_{\alf\beta} + R^{\alf\beta\gamma\delta}R_{\alf\beta\gamma\delta}
\r)}_{l_{(2)}}]
 \m{(18)}
\ee
where $M_*$ is the string mass scale, $\alf \propto M_*^{-2}$
and $l_{(2)}$ is the Gauss-Bonnet term.
Like in Einstein's theory
 \cite{[1]}, we present
a perturbed (with respect to a background spacetime)
scenario for (\ref{(18)}) by the Lagrangian:
\be
\lag^{(5)}_{pert} = \lag^{(5)} - \Bar {\lag^{(5)}}
+ div.
\m{(19)}
\ee

To construct a superpotential of the type (11) corresponding to
(\ref{(19)}) one has to go as follows.
Keeping in mind (\ref{(17)}) we present
$\lag^{(5)}$   in the form (\ref{(1)}) and substitute it into (\ref{(11)}).
To obtain a contribution from  $\Bar {\lag^{(5)}}$
we do the ``bar'' operation under the previous result. Finally,
to take into account  $div$ in (\ref{(19)}) we use
(\ref{(12)}).
Here, we do not present an expression for this
superpotential  explicitly, it coincides with the one given recently
in \cite{D-Katz-O}. We note  only that
a choice of a divergence in (\ref{(19)})
is in  a one-to-one correspondance with
boundary conditions under variation, and for each of divergences an
unique superpotential corresponds. The Dirichlet
boundary conditions  are also
permissible.

Let us turn to the new superpotential (\ref{(16)})
that does not depend on divergences in the Lagrangian.
The coefficient (5) corresponding $\hat l_{(2)} \equiv \sqrt{-g} l_{(2)}$
in (\ref{(18)}) with presentation (\ref{(17)}) is
$$
\hat n^{\alf\beta\gamma}_{(5)\sig} =
\l(g^{\alf(\beta}\delta^{\gamma)}_\rho - \delta^\alf_\rho g^{\beta\gamma}\r)
\l(\hat R \delta^\rho_\sig - 8 \hat R^\rho_\sig\r)
+ 4 \hat R^{\alf(\beta\gamma)}{}_\sig.
$$
Then the superpotential (\ref{(16)}), corresponding
(\ref{(19)}), acquires the form
\bea
 \hat
 \Phi^{\alf\beta}_{(5b)} & \equiv & M^3_*\l[
 \delta\hat g^{\sig[\alf} \xi^{\beta]}{}_{;\sig}
 +
\xi^{[\alf} \delta\hat g^{\beta]\sig}{}_{;\sig}-
   \Bar g^{\rho[\alf}
\delta \hat g^{\beta]\sig}{}_{;\rho} \xi_\sig
\r] \nonumber \\
&{}& +~  \alf M^3_* \l[{\textstyle{2\over 3}} \delta\hat
 n^{[\alf\beta]\lam}_{(5)\sig}\xi^\sig{}_{;\lam} -
\l({\textstyle{1\over 3}} \delta\hat n^{[\alf\beta]\rho}_{(5)\sig~;\rho}+
\delta\hat
 n^{\tau\rho[\alf}_{(5)\lam~;\tau} \bar g^{\beta]\lam} \bar g_{\rho\sig}\r)
\xi^\sig \r]
 \m{(20)}
\eea
 with  $\hat g^{\alf\beta} \equiv \sqrt{-g}g^{\alf\beta}$, $\delta \hat g^{\alf\beta} =
 \hat g^{\alf\beta} - \Bar{\hat g^{\alf\beta}}$ and
  $\delta \hat n^{\alf\beta\gamma}_{(5)\sig} =
 \hat n^{\alf\beta\gamma}_{(5)\sig} -
 \Bar {\hat n^{\alf\beta\gamma}_{(5)\sig}}$.

The present work has some connections with previous results
in \cite{[1]} and \cite{[7]}.
In the work by Katz, Bi\v c\'ak and Lynden-Bell \cite{[1]} (KBL),
with using the standard
procedure (analogously to (\ref{(1)}) - (\ref{(12)}) here)
superpotentials and currents for perturbations
on arbitrary curved backgrounds
in GR were constructed.
In our work \cite{[7]},
the KBL system was modified by the Belinfante method.

The  Lagrangian (\ref{(19)}) transfers to the KBL Lagrangian \cite{[1]}:
$\lag^{(5)}_{pert} \goto \lag_G$ after changing $D=5$ by $D=4$, and
if $M_*^{-3} = \kappa$
(Einstein constant),  $\alf =0$ and $ div = (2\kappa)^{-2}(\hat g^{\rho\sig}
\Delta^\nu_{\rho\sig} - \hat
g^{\nu\rho}\Delta^\sig_{\rho\sig})_{,\nu} $.  The substitution of $\lag_G$
into (\ref{(7)}), (\ref{(11)}) and (\ref{(12)})
gives exactly the KBL  conservation laws \cite{[1]}. Our
conclusion is, thus, that the KBL results in the sense of the
procedure are uniquely defined by the KBL Lagrangian $\lag_G$.
The problem of uniqueness in definition of the KBL currents and
superpotentials already was considered.  So, Julia and Silva \cite{[2],
[3]}, and independently Chen and Nester \cite{[4]}, stated that the KBL
quantities are {\it unique} defined by Dirichlet boundary
conditions. Thus, our results are
in a correspondance with \cite{[2],[3],[4]},
Besides, changing
the  divergence in the Lagrangian  $\lag_G$ we define {\it uniquelly}
another superpotentials and currents with different properties.
An analogous assertion was done in \cite{[13]} in the framework of
the covariant Hamiltonian approach.

Concerning our results in \cite{[7]}, substitution of the KBL
Lagrangian $\lag_G$ into Eqs. (\ref{(13)}) - (\ref{(16)}) gives exactly
corresponding  expressions for GR obtained by the Belinfante method
\cite{[7]}.  By this we  conclude  that the  conservation laws in
\cite{[7]}
are {\it uniquely} defined by $\lag_G$ in the sense of the
procedure. Note also, that the superpotential (\ref{(20)}) becomes
the
one in \cite{[7]} for $D=4$  setting
$M_*^{-3} = \kappa$ and  $\alf =0$.

\medskip

\noindent{\bf  Acknowledgments.}
The author is  grateful to Joseph Katz  for
very useful and important remarks and discussions, for advices
that helped to formulate the paper in the present form.
He thanks  Laszlo Szabados for explanation of their works and helpful
conversations. He expresses also his gratitudes to Piotr
Chru\'sciel the discussions with whom at the beginning stage
initiated the present consideration.

\ed